\documentclass[prd,aps,showpacs,epsf,floats,twocolumn]{revtex4}%
\usepackage{amsfonts}
\usepackage{amsmath}
\usepackage{amssymb}
\usepackage{graphicx}%
\providecommand{\U}[1]{\protect\rule{.1in}{.1in}}
\begin{document}
\title{\textbf{Information geometric complexity of entropic motion on curved
statistical manifolds\ }}
\author{Carlo Cafaro$^{1,2}$}
\affiliation{$^{1}$Max-Planck Institute for the Science of Light, 91058 Erlangen, Germany }
\affiliation{$^{2}$Institute of Physics, Johannes Gutenberg University Mainz, 55128 Mainz, Germany}

\begin{abstract}
Physical systems behave according to their underlying dynamical equations
which, in turn, can be identified from experimental data. Explaining data
requires selecting mathematical models that best capture the data
regularities. Identifying dynamical equations from the available data and
statistical model selection are both very difficult tasks.

Motivated by these fundamental links among physical systems, dynamical
equations, experimental data and statistical modeling, we discuss in this
invited Contribution our information geometric measure of complexity of
geodesic paths on curved statistical manifolds\ underlying the entropic
dynamics of classical physical systems described by probability distributions.
We also provide several illustrative examples of entropic dynamical models
used to infer macroscopic predictions when only partial knowledge of the
microscopic nature of the system is available. Finally, we present entropic
arguments to briefly address complexity softening effects due to statistical
embedding procedures.

\end{abstract}

\maketitle

\section{Introduction}

The intimate connection between dynamics, on the one hand, and modeling,
prediction, and complexity, on the other, is quite remarkable in science
\cite{jimmy}. In real-world experiments, we usually gather data of the state
of a physical system at various points in space and time. Then, to achieve
some comprehension of the physics behind the behaviour of the system, we must
reconstruct the underlying dynamical equations from the data. Deducing
dynamics from experimental observations (data) is a fundamental part of
science \cite{minghia1, minghia2}. We observe the trajectories of planets to
deduce the laws of celestial mechanics; we consider monetary parameters to
determine economic laws; we observe atoms to deduce quantum mechanics. A
current challenge is the analysis of data gathered from networks of
interferometric gravitational-wave detectors to search for a stochastic
gravitational-wave background \cite{cella}.

A very recent and extremely interesting work shows that deducing the
underlying dynamical equations from experimental data is NP hard (the NP
complexity class denotes a class of problems that have solutions which can be
quickly checked on a classical computer) and is computationally intractable
\cite{toby}. This hardness result holds true for both classical and quantum
systems, and regardless of how much experimental data we gather about the
system. These results imply that various closely related problems, such as
finding the dynamical equation that best approximates the data, or testing a
dynamical model against experimental data, are intractable in general.

By\textbf{\ }analyzing the available data about a system of interest, it is
possible to identify classes of regularities of the system itself. It is
generally agreed that something almost entirely random, with practically no
regularities, would have an effective complexity near zero \cite{gell-mann}.
Instead, structured systems (where correlations among system's constituents
arise) can be very complex. Structure and correlation are not completely
independent of randomness. Indeed, both maximally random and
perfectly\textbf{\ }ordered systems possess no structure \cite{james1,
james2}. What then is the meaning of complexity? It appears that:

\begin{itemize}
\item A good measure of complexity is best justified through utility in
further application \cite{landauer};

\item A good measure of complexity is most useful for comparison between
things, at least one of which, has high complexity by that measure
\cite{gell-mann};

\item A good measure of complexity for many-body systems ought to obey the
so-called slow law growth \cite{bennett-book}: complexity ought not to
increase quickly, except with low probability, but can increase slowly;

\item A good measure of complexity is one for which the motivations for its
introduction and the features it is intended to capture are stated in a clear
manner \cite{james1}.
\end{itemize}

In general, good measures of complexity are introduced within formulations
that deal with the whole sequence of events that lead to the object whose
complexity is being described \cite{landauer}. For such measures, that which
is reached only through a difficult path is complex. For instance, when
defining the complexity of a noisy quantum channel, the concept of pattern
plays a role, in some sense \cite{romano}. The thermodynamic and the logical
depths are two such measures as well. The thermodynamic depth is the measure
of complexity proposed by Lloyd and Pagels and it represents the amount of
entropy produced during a state's actual evolution \cite{lloyd1}. The logical
depth is a measure of complexity proposed by Bennett and it represents the
execution time required for a universal Turing machine to run the minimal
program that reproduces (say) a system's configuration \cite{bennett-f}.

Since the path leading to an object (or, state) is central when defining a
measure of complexity, simple thermodynamic criteria applied to the states to
be compared are inadequate. Thermodynamic potentials measure a system's
capacity for irreversible change, but do not agree with intuitive notions of
complexity \cite{bennett-book}. For instance, the thermodynamic entropy, a
measure of randomness, is a monotonic function of temperature where high (low)
temperature corresponds to high (low) randomness. However, given that there
are many functions that vanish in the extreme ordered and disordered limits,
it is clear that requiring this property does not sufficiently restrict a
complexity measure of statistical nature (statistical complexity \cite{james2}
is a quantity that measures the amount of memory needed, on average, to
statistically reproduce a given configuration). Despite these facts, it is
undisputable that thermodynamics does play a key role when investigating
qualitative differences in the complexity of reversible and dissipative
systems \cite{bennett-f}.

The difficulty of constructing a good theory from a data set can be roughly
identified with cripticity while the difficulty of making predictions from the
theory can be regarded as a rough interpretation of logical depth. Both
cripticity and logical depth are intimately related to the concept of
complexity. Making predictions can be very difficult in general, especially in
composite systems where interactions between subsystems are introduced. The
introduction of interactions leads to fluctuation growth which, in turn, can
cause the dynamics to become nonlinear and chaotic. Such phenomena are very
common and can occur in both natural (cluster of stars) and artificial
(financial network) complex dynamical systems \cite{lloyd2013}. A fundamental
problem in the physics of complex systems is model reduction, that is finding
a low-dimensional model that captures the gross features of a high-dimensional
system \cite{jie}. Sometimes, to make reliable macroscopic predictions,
considering the dynamics alone may not be sufficient and entropic arguments
should be taken into account as well \cite{jaynes1}.

As\textbf{\ }stated earlier, one of the major goals of physics is modelling
and predicting natural phenomena using relevant information about the system
of interest. Taking this statement seriously, it is reasonable to expect that
the laws of physics should reflect the methods for manipulating information.
This point of view constitutes quite a departure from the conventional line of
thinking where laws of physics are used to manipulate information. For
instance, in quantum information science, information is manipulated using the
laws of quantum mechanics. This alternative perspective is best represented in
the so-called Entropic Dynamics (ED) \cite{caticha-ED}, a theoretical
framework built on both maximum relative entropy (MrE) methods
\cite{caticha-giffin} and information geometric techniques \cite{amari-japan}.
The most intriguing question being pursued in ED stems from the possibility of
deriving dynamics from purely entropic arguments. Indeed, the ED approach has
already been applied for the derivation of Newton's dynamics
\cite{caticha-cafaro} and aspects of quantum theory \cite{caticha-jpa}.

In this invited Contribution, inspired by the ED approach to physics and
motivated by these fundamental links among physical systems, dynamical
equations, experimental data and statistical modeling, we present our
information geometric measure of complexity of geodesic paths on curved
statistical manifolds underlying the entropic dynamics of classical physical
systems described by probability distributions. We also provide several
illustrative examples of entropic dynamical models used to infer macroscopic
predictions when only partial knowledge of the microscopic nature of the
system is available. Finally, we emphasize the relevance of entropic arguments
in addressing complexity softening effects due to statistical embedding procedures.

\section{Complexity}

In \cite{carlo-tesi}, the so-called Information Geometric Approach to Chaos
(IGAC) was presented. \ The IGAC uses the ED formalism to study the complexity
of informational geodesic flows on curved statistical manifolds underlying the
entropic dynamics of classical physical systems described by probability distributions.

A geodesic on a curved statistical manifold $\mathcal{M}_{S}$ represents the
maximum probability path a complex dynamical system explores in its evolution
between initial and final macrostates. Each point of the geodesic is
parametrized by the macroscopic dynamical variables\textbf{\ }$\left\{
\theta\right\}  $ defining the macrostate of the system. Furthermore, each
macrostate is in a one-to-one correspondence with the probability distribution
$\left\{  p\left(  x|\theta\right)  \right\}  $ representing the maximally
probable description of the system being considered. The quantity $x$ is a
microstate of the microspace $\mathcal{X}$. The set of macrostates forms the
parameter space $\mathcal{D}_{\theta}$ while the set of probability
distributions forms the statistical manifold $\mathcal{M}_{S}$.

The IGAC\ is the information geometric analogue of conventional
geometrodynamical approaches \cite{casetti, di bari} where the classical
configuration space\ is being replaced by a statistical manifold with the
additional possibility of considering chaotic dynamics arising from non
conformally flat metrics (the Jacobi metric is always conformally flat,
instead). It is an information geometric extension of the Jacobi
geometrodynamics (the geometrization of a Hamiltonian system by transforming
it to a geodesic flow \cite{jacobi}).

The reformulation of dynamics in terms of a geodesic problem allows the
application of a wide range of well-known geometrical techniques in the
investigation of the solution space and properties of the equation of motion.
The power of the Jacobi reformulation is that all of the dynamical information
is collected into a single geometric object in which all the available
manifest symmetries are retained- the manifold on which geodesic flow is
induced. For example, integrability of the system is connected with existence
of Killing vectors and tensors on this manifold. The sensitive dependence of
trajectories on initial conditions, which is a key ingredient of chaos, can be
investigated from the equation of geodesic deviation. In the Riemannian
\cite{casetti} and Finslerian \cite{di bari} (a Finsler metric is obtained
from a Riemannian metric by relaxing the requirement that the metric be
quadratic on each tangent space) geometrodynamical approach to chaos in
classical Hamiltonian systems, a very challenging problem is finding a
rigorous relation among sectional curvatures, Lyapunov exponents, and the
Kolmogorov-Sinai dynamical entropy \cite{kawabe}.

\subsection{Information metric}

An $n$-dimensional $%
\mathbb{C}
^{\infty}$ differentiable manifold is a set of points $\mathcal{M}$ admitting
coordinate systems $\mathcal{C}_{\mathcal{M}}$ and satisfies the following two
conditions: 1) each element $c\in\mathcal{C}_{\mathcal{M}}$ is a one-to-one
mapping from $\mathcal{M}$ to some open subset of $%
\mathbb{R}
^{n}$; 2) For all $c\in\mathcal{C}_{\mathcal{M}}$, given any one-to-one
mapping $\xi$ from $\mathcal{M}$ to $%
\mathbb{R}
^{n}$, we have that $\xi\in\mathcal{C}_{\mathcal{M}}\Leftrightarrow\xi\circ
c^{-1}$ is a $%
\mathbb{C}
^{\infty}$ diffeomorphism. In this Contribution, the points of $\mathcal{M}$
are probability distributions. Furthermore, we consider Riemannian manifolds
$\left(  \mathcal{M}\text{, }g\right)  $. The Riemannian metric $g$ is not
naturally determined by the structure of $\mathcal{M}$ as a manifold. In
principle, it is possible to consider an infinite number of Riemannian metrics
on $\mathcal{M}$. A fundamental assumption in the information geometric
framework is the choice of the Fisher-Rao information metric as the metric
that underlies the Riemannian geometry of probability distributions
\cite{amari-japan, fisher, rao}, namely%
\begin{equation}
g_{\mu\nu}\left(  \theta\right)  \overset{\text{def}}{=}\int dxp\left(
x|\theta\right)  \partial_{\mu}\log p\left(  x|\theta\right)  \partial_{\nu
}\log p\left(  x|\theta\right)  \text{,} \label{FR}%
\end{equation}
with $\mu$, $\nu=1$,..., $n$ for an $n$-dimensional manifold and
$\partial_{\mu}\overset{\text{def}}{=}\frac{\partial}{\partial\theta^{\mu}}$.
The quantity $x$ labels the microstates of the system. The choice of the
information metric can be motivated in several ways, the strongest of which is
Cencov's characterization theorem \cite{cencov}. In this theorem, Cencov
proves that the information metric is the only Riemannian metric (except for a
constant scale factor) that is invariant under a family of probabilistically
meaningful mappings termed congruent embeddings by Markov morphism
\cite{cencov, campbell}.

Given a statistical manifold $\mathcal{M}_{S}$ with a metric $g_{\mu\nu}$, the
ED is concerned with the following issue \cite{caticha-ED}: given the initial
and final states, what trajectory is the system expected to follow? The answer
turns out to be that the expected trajectory is the geodesic that passes
through the given initial and final states. Furthermore, the trajectory
follows from a principle of inference, the MrE method \cite{caticha-giffin}.
The objective of the MrE method is to update from a prior distribution $q$\ to
a posterior distribution $P(x)$\ given the information that the posterior lies
within a certain family of distributions $p$. The selected posterior
$P(x)$\ is that which maximizes the logarithm relative entropy $\mathcal{S}%
[p\left\vert q\right.  ]$,%
\begin{equation}
\mathcal{S}[p\left\vert q\right.  ]\overset{\text{def}}{=}-%
{\displaystyle\int}
dxp\left(  x\right)  \log\frac{p\left(  x\right)  }{q\left(  x\right)
}\text{.}%
\end{equation}
Since prior information is valuable, the functional $\mathcal{S}[p\left\vert
q\right.  ]$\ has been chosen so that rational beliefs are updated only to the
extent required by the new information. We emphasize that ED is formally
similar to other generally covariant theories: the dynamics is reversible, the
trajectories are geodesics, the system supplies its own notion of an intrinsic
time, the motion can be derived from a variational principle of the form of
Jacobi's action principle rather than the more familiar principle of Hamilton.
In short, the canonical Hamiltonian formulation of ED is an example of a
constrained information-dynamics where the information-constraints play the
role of generators of evolution. For more details on the ED, we refer to
\cite{caticha-ED}.

A geodesic on a $n$-dimensional curved statistical manifold $\mathcal{M}_{S}$
represents the maximum probability path a complex dynamical system explores in
its evolution between initial and final macrostates $\theta_{\text{initial}}$
and $\theta_{\text{final}}$, respectively. Each point of the geodesic
represents a macrostate parametrized by the macroscopic dynamical variables
$\theta\equiv\left(  \theta^{1}\text{,..., }\theta^{n}\right)  $ defining the
macrostate of the system. Each component $\theta^{k}$ with $k=1$,..., $n$ is a
solution of the geodesic equation \cite{caticha-ED},%
\begin{equation}
\frac{d^{2}\theta^{k}}{d\tau^{2}}+\Gamma_{lm}^{k}\frac{d\theta^{l}}{d\tau
}\frac{d\theta^{m}}{d\tau}=0\text{.}%
\end{equation}
Furthermore, as stated earlier, each macrostate $\theta$ is in a one-to-one
correspondence with the probability distribution $p\left(  x|\theta\right)  $.
This is a distribution of the microstates $x$.

\subsection{Entropic motion}

The main objective of ED\ is to derive the expected trajectory of a system,
assuming it evolves from a known initial state $\theta_{i}$ to a known final
state $\theta_{f}$. The ED framework implicitly assumes there exists a
trajectory, in the sense that, large changes are the result of a continuous
succession of very many small changes. Therefore, the problem of studying
large changes is reduced to the much simpler problem of studying small
changes. Focusing on small changes and assuming that the change in going from
the initial state $\theta_{i}$ to the final state $\theta_{f}=\theta
_{i}+\Delta\theta$ is sufficiently small, the distance $\Delta l$ between such
states becomes,%
\begin{equation}
\Delta l^{2}\overset{\text{def}}{=}g_{\mu\nu}\left(  \theta\right)
\Delta\theta^{\mu}\Delta\theta^{\nu}\text{.}%
\end{equation}
Following Caticha's work in \cite{caticha-ED}, we explain how to determine
which states are expected to lie on the expected trajectory between
$\theta_{i}$ and $\theta_{f}$. First, in going from the initial to the final
state the system must pass through a halfway point, that is, a state $\theta$
that is equidistant from $\theta_{i}$ and $\theta_{f}$. Upon choosing\textbf{
}the halfway state, the expected trajectory of the system can be determined.
Indeed, there is nothing special about halfway states. For instance, we could
have similarly argued that in going from the initial to the final state the
system must first traverse a third of the way, that is, it must pass through a
state that is twice as distant from $\theta_{f}$ as it is from $\theta_{i}$.
In general, the system must pass through an intermediate states $\theta_{\xi}$
such that, having already moved a distance $dl$ away from the initial
$\theta_{i}$, there remains a distance $\xi dl$ to be covered to reach the
final $\theta_{f}$ . Halfway states have $\xi$ $=1$, third of the way states
have $\xi$ $=2$, and so on. Each different value of $\xi$ provides a different
criterion to select the trajectory. If there are several ways to determine a
trajectory, consistency requires that all these ways should agree. The
selected trajectory must be independent of $\xi$. Therefore, the main ED
problem becomes the following: initially, the system is in state
$p(x|\theta_{i})$ and new information in the form of constraints is given to
us; the system has moved to one of the neighboring states in the family
$p(x|\theta_{\xi})$; the problem becomes that of selecting the proper
$p(x|\theta_{\xi})$. This new formulation of the ED problem is precisely the
kind of problem to be addressed using the MrE method. We recall that the MrE
method is a method for processing information. It allows us to go from an old
set of rational beliefs, described by the prior probability distribution, to a
new set of rational beliefs, described by the posterior distribution, when the
available information is just a specification of the family of distributions
from which the posterior must be selected. Usually, this family of posteriors
is defined by the known expected values of some relevant variables. It should
be noted however, that it is not strictly necessary for the family of
posteriors to be defined via expectation values, nor does the
information-constraints need to be linear functionals. In ED, constraints are
defined geometrically. Whenever one contemplates using the MrE method, it is
important to specify which entropy should be maximized. The selection of a
distribution $p(x|\theta)$ requires that the entropies to be considered must
be of the form,%
\begin{equation}
S\left[  p|q\right]  \overset{\text{def}}{=}-\int dxp\left(  x|\theta\right)
\log\left(  \frac{p\left(  x|\theta\right)  }{q\left(  x\right)  }\right)
\text{.} \label{juve}%
\end{equation}
Equation (\ref{juve}) defines the entropy of $p\left(  x|\theta\right)  $
relative to the prior $q(x)$. The interpretation of $q(x)$ as the prior
follows from the logic behind the MrE method itself. The selected posterior
distribution should coincide with the prior distribution when there are no
constraints. Since the distribution that maximizes $S\left[  p|q\right]  $
subject to no constraints is $p\propto q$, we must set $q(x)$ equal to the
prior. That said, let us return to our ED problem. Assuming we know that the
system is initially in state $p(x|\theta_{i})$ but have obtained no
information\textbf{ }reflecting that the system has moved. We therefore have
no reason to believe that any change has occurred. The prior $q(x)$ should be
chosen so that the maximization of $S\left[  p|q\right]  $ subject to no
constraints leads to the posterior $p=$ $p(x|\theta_{i})$. The correct choice
is $q(x)=$ $p(x|\theta_{i})$. If on the other hand we know that the system is
initially in state $p(x|\theta_{i})$ and furthermore, we obtain information
that the system has moved to one of the neighboring states in the family
$p(x|\theta_{\xi})$, then the correct selection of the posterior probability
distribution is obtained by maximizing the entropy,%
\begin{equation}
S\left[  \theta|\theta_{i}\right]  \overset{\text{def}}{=}-\int dxp(x|\theta
)\log\left(  \frac{p(x|\theta)}{p(x|\theta_{i})}\right)  \text{,}%
\end{equation}
subject to the constraint $\theta=\theta_{\xi}$. For the sake of reasoning,
let us assume that the system evolves from a known initial state $\theta_{i}$
to a known final state $\theta_{f}=\theta_{i}+\Delta\theta$. Furthermore, let
us denote with $\theta_{\xi}$ $=$ $\theta_{i}$ $+d\theta$ ($\xi\in%
\mathbb{R}
_{0}^{+}$) an arbitrary intermediate state infinitesimally close to
$\theta_{i}$. Thus, the distance $d\left(  \theta_{i}\text{, }\theta
_{f}\right)  \overset{\text{def}}{=}dl_{i\rightarrow f}^{2}$ between
$\theta_{i}$ to and $\theta_{f}$ is given by,%
\begin{equation}
dl_{i\rightarrow f}^{2}\overset{\text{def}}{=}g_{\mu\nu}\left(  \theta\right)
\Delta\theta^{\mu}\Delta\theta^{\nu}\text{,}%
\end{equation}
while the distance between $\theta_{i}$ to and $\theta_{\xi}$ reads,%
\begin{equation}
dl_{i\rightarrow\xi}^{2}\overset{\text{def}}{=}g_{\mu\nu}\left(
\theta\right)  d\theta^{\mu}d\theta^{\nu}\text{.} \label{1}%
\end{equation}
Finally, the distance between $\theta_{\xi}$ and $\theta_{f}$ becomes,%
\begin{equation}
dl_{\xi\rightarrow f}^{2}\overset{\text{def}}{=}g_{\mu\nu}\left(
\theta\right)  \left(  \Delta\theta^{\mu}-d\theta^{\mu}\right)  \left(
\Delta\theta^{\nu}-d\theta^{\nu}\right)  \text{.} \label{2}%
\end{equation}
The MrE\ maximization problem is to maximize $S[\theta_{\xi}|\theta_{i}]=$
$S\left[  \theta_{i}+d\theta|\theta_{i}\right]  $,%
\begin{equation}
S\left[  \theta_{i}+d\theta|\theta_{i}\right]  \overset{\text{def}}{=}%
-\frac{1}{2}g_{\mu\nu}\left(  \theta\right)  d\theta^{\mu}d\theta^{\nu}%
=-\frac{1}{2}dl_{i\rightarrow\xi}^{2}\text{,}%
\end{equation}
under variations of $d\theta$ subject to the geometric constraint,%
\begin{equation}
\xi dl_{i\rightarrow\xi}=dl_{\xi\rightarrow f}\text{,} \label{constraint}%
\end{equation}
or equivalently, $\xi^{2}dl_{i\rightarrow\xi}^{2}-dl_{\xi\rightarrow f}^{2}%
=0$. It must then be true that,%
\begin{equation}
\delta\left[  -\frac{1}{2}g_{\mu\nu}\left(  \theta\right)  d\theta^{\mu
}d\theta^{\nu}-\lambda\left(  \xi^{2}dl_{i\rightarrow\xi}^{2}-dl_{\xi
\rightarrow f}^{2}\right)  \right]  =0\text{,} \label{3}%
\end{equation}
where $\lambda$ denotes a Lagrangian multiplier. Substituting Eqs. (\ref{1})
and (\ref{2}) into Eq. (\ref{3}), we obtain%
\begin{equation}
\left\{  \left[  1+2\lambda\left(  \xi^{2}-1\right)  \right]  d\theta_{\mu
}+2\lambda\Delta\theta_{\mu}\right\}  \delta\left(  d\theta^{\mu}\right)
=0\text{.} \label{4}%
\end{equation}
Since (\ref{4}) must hold for any $\delta\left(  d\theta^{\mu}\right)  $, it
must be\textbf{ }the case that%
\begin{equation}
\left\{  \left[  1+2\lambda\left(  \xi^{2}-1\right)  \right]  d\theta_{\mu
}+2\lambda\Delta\theta_{\mu}\right\}  =0\text{,}%
\end{equation}
that is,%
\begin{equation}
d\theta_{\mu}=\chi\Delta\theta_{\mu}\text{,} \label{5}%
\end{equation}
where $\chi=\chi\left(  \xi\text{, }\lambda\right)  $ is defined as,%
\begin{equation}
\chi\left(  \xi\text{, }\lambda\right)  \overset{\text{def}}{=}\frac
{1}{\left(  1-\xi^{2}\right)  -\frac{1}{2\lambda}}\text{.} \label{6}%
\end{equation}
To find the value of the Lagrange multiplier $\lambda$, observe that the
geometric constraint in Eq. (\ref{constraint}) can be rewritten as, $\xi
^{2}dl_{i\rightarrow\xi}^{2}-dl_{\xi\rightarrow f}^{2}=0$. Then, using Eqs.
(\ref{1}), (\ref{2}) and (\ref{5}), we obtain%
\begin{equation}
\left[  \xi^{2}\chi^{2}-\left(  1-\chi\right)  ^{2}\right]  g_{\mu\nu}\left(
\theta\right)  \Delta\theta^{\mu}\Delta\theta^{\nu}=0\text{,}%
\end{equation}
thus,%
\begin{equation}
\xi^{2}\chi^{2}-\left(  1-\chi\right)  ^{2}=0\text{.} \label{7}%
\end{equation}
Combining Eqs. (\ref{6}) and (\ref{7}), we find%
\begin{equation}
\chi\left(  \xi\right)  \overset{\text{def}}{=}\frac{1}{1+\xi}\text{ and,
}\lambda\left(  \xi\right)  \overset{\text{def}}{=}-\frac{1}{2\xi\left(
1+\xi\right)  }\text{.}%
\end{equation}
In conclusion, it has been determined that%
\begin{equation}
dl_{i\rightarrow\xi}^{2}\overset{\text{def}}{=}\frac{1}{\left(  1+\xi\right)
^{2}}\Delta\theta^{2}\text{,} \label{8}%
\end{equation}
and,%
\begin{equation}
dl_{\xi\rightarrow f}^{2}\overset{\text{def}}{=}\frac{\xi^{2}}{\left(
1+\xi\right)  ^{2}}\Delta\theta^{2}\text{.} \label{9}%
\end{equation}
From Eqs. (\ref{8}) and (\ref{9}), it follows that%
\begin{equation}
dl_{i\rightarrow\xi}+dl_{\xi\rightarrow f}=\frac{1}{1+\xi}\Delta\theta\text{
}+\frac{\xi}{1+\xi}\Delta\theta=\Delta\theta\text{.} \label{10}%
\end{equation}
However, recall that $dl_{i\rightarrow f}^{2}\overset{\text{def}}{=}g_{\mu\nu
}\left(  \theta\right)  \Delta\theta^{\mu}\Delta\theta^{\nu}=\Delta\theta^{2}%
$, that is%
\begin{equation}
dl_{i\rightarrow f}=\Delta\theta\text{.} \label{11}%
\end{equation}
Combining Eqs. (\ref{10}) and (\ref{11}), we arrive at%
\begin{equation}
dl_{i\rightarrow f}=dl_{i\rightarrow\xi}+dl_{\xi\rightarrow f}\text{.}%
\end{equation}
In other words, given
\begin{equation}
\Delta\theta\overset{\text{def}}{=}d\theta+\left(  \Delta\theta-d\theta
\right)  \text{,} \label{12}%
\end{equation}
we have shown by means of entropic arguments that,%
\begin{equation}
\left\Vert \Delta\theta\right\Vert =\left\Vert d\theta\right\Vert +\left\Vert
\Delta\theta-d\theta\right\Vert \text{,} \label{13}%
\end{equation}
where $\left\Vert \Delta\theta\right\Vert \overset{\text{def}}{=}%
\sqrt{dl_{i\rightarrow f}^{2}}$, $\left\Vert d\theta\right\Vert \overset
{\text{def}}{=}\sqrt{dl_{i\rightarrow\xi}^{2}}$ and, $\left\Vert \Delta
\theta-d\theta\right\Vert \overset{\text{def}}{=}\sqrt{dl_{\xi\rightarrow
f}^{2}}$. Given Eq. (\ref{12}), Eq. (\ref{13}) holds true iff $d\theta$ and
$\Delta\theta-d\theta$ are collinear. Therefore, the expected trajectory is a
straight line: the triangle defined by the points $\theta_{i}$, $\theta_{\xi}%
$, and $\theta_{f}$ degenerates into a straight line. This is sufficient to
determine a short segment of the trajectory: all intermediate states lie on
the straight line between $\theta_{i}$ and $\theta_{f}$. The generalization
beyond short trajectories is immediate: if any three nearby points along a
curve lie on a straight line the curve is a \emph{geodesic}. This result is
independent of the arbitrarily chosen value $\xi$ so the potential consistency
problem we mentioned before does not arise. Summarizing, the answer to the ED
problem is the following: \emph{the expected trajectory between a known
initial and final state is the geodesic that passes through them.} However,
the question of whether the actual trajectory is the expected trajectory
remains unanswered and depends on whether the information encoded in the
initial state is sufficient for prediction.

\subsection{Volumes in curved statistical manifolds}

Once the distances among probability distributions have been assigned using
the Fisher-Rao information metric tensor $g_{\mu\nu}\left(  \theta\right)  $,
a natural next step is to obtain measures for extended regions in the space of
distributions. Consider an $n$-dimensional volume of the statistical manifold
$\mathcal{M}_{s}$ of distributions $p\left(  x|\theta\right)  $ labelled by
parameters $\theta^{\mu}$ with $\mu=1$,..., $n$. The parameters $\theta^{\mu}$
are coordinates for the point $p$ and in these coordinates it may not be
obvious how to write an expression for a volume element $d\mathcal{V}%
_{\mathcal{M}_{s}}$. However, within a sufficiently small region any curved
space looks flat. That is to say, curved spaces are locally flat. The idea
then is rather simple: within that very small region, we should use Cartesian
coordinates wherein the metric takes a very simple form, namely the identity
matrix $\delta_{\mu\nu}$. In locally Cartesian coordinates $\chi^{\alpha}$ the
volume element is given by the product $d\mathcal{V}_{\mathcal{M}_{s}}%
\overset{\text{def}}{=}d\chi^{1}d\chi^{2}$.....$d\chi^{n}$, which in terms of
the old coordinates reads,%
\begin{equation}
d\mathcal{V}_{\mathcal{M}_{s}}\overset{\text{def}}{=}\left\vert \frac
{\partial\chi}{\partial\theta}\right\vert d\theta^{1}d\theta^{2}\text{...
}d\theta^{n}\text{.}%
\end{equation}
The problem at hand then is the calculation of the Jacobian $\left\vert
\frac{\partial\chi}{\partial\theta}\right\vert $ of the transformation that
takes the metric $g_{\mu\nu}$ into its Euclidean form $\delta_{\mu\nu}$. Let
the new coordinates be defined by $\chi^{\mu}\overset{\text{def}}{=}\Xi^{\mu
}\left(  \theta^{1}\text{,...., }\theta^{n}\right)  $ where $\Xi$ denotes a
coordinates transformation map. A small change $d\theta$ corresponds to a
small change $d\chi$,%
\begin{equation}
d\chi^{\mu}\overset{\text{def}}{=}X_{m}^{\mu}d\theta^{m}\text{ where }%
X_{m}^{\mu}\overset{\text{def}}{=}\frac{\partial\chi^{\mu}}{\partial\theta
^{m}}\text{,}%
\end{equation}
and the Jacobian is given by the determinant of the matrix $X_{m}^{\mu}$,
$\left\vert \frac{\partial\chi}{\partial\theta}\right\vert \overset
{\text{def}}{=}\left\vert \det\left(  X_{m}^{\mu}\right)  \right\vert $. The
distance between two neighboring points is the same whether we compute it in
terms of the old or the new coordinates, $dl^{2}=g_{\mu\nu}d\theta^{\mu
}d\theta^{\nu}=\delta_{\alpha\beta}d\chi^{\alpha}d\chi^{\beta}$. Therefore the
relation between the old and the new metric is $g_{\mu\nu}=\delta_{\alpha
\beta}X_{\mu}^{\alpha}X_{\nu}^{\beta}$. Taking the determinant of $g_{\mu\nu}%
$, we obtain $g\overset{\text{def}}{=}\det\left(  g_{\mu\nu}\right)  =\left[
\det\left(  X_{\mu}^{\alpha}\right)  \right]  ^{2}$ and therefore $\left\vert
\det\left(  X_{\mu}^{\alpha}\right)  \right\vert =\sqrt{g}$. Finally, we have
succeeded in expressing the volume element totally in terms of the coordinates
$\theta$ and the known metric $g_{\mu\nu}\left(  \theta\right)  $,
$dV_{\mathcal{M}_{s}}\overset{\text{def}}{=}\sqrt{g}d^{n}\theta$. Thus, the
volume of any extended region on the manifold is given by,%
\begin{equation}
\mathcal{V}_{\mathcal{M}_{s}}\overset{\text{def}}{=}\int d\mathcal{V}%
_{\mathcal{M}_{s}}=\int\sqrt{g}d^{n}\theta\text{.}%
\end{equation}
Observe that $\sqrt{g}d^{n}\theta$ is a scalar quantity and is therefore
invariant under orientation preserving general coordinate transformations
$\theta\rightarrow\theta^{\prime}$. The square root of the determinant
$g\left(  \theta\right)  $ of the metric tensor $g_{\mu\nu}\left(
\theta\right)  $ and the flat infinitesimal volume element $d^{n}\theta$
transform as,%
\begin{equation}
\sqrt{g\left(  \theta\right)  }\overset{\theta\rightarrow\theta^{\prime}%
}{\rightarrow}\left\vert \frac{\partial\theta^{\prime}}{\partial\theta
}\right\vert \sqrt{g\left(  \theta^{\prime}\right)  }\text{, }d^{n}%
\theta\overset{\theta\rightarrow\theta^{\prime}}{\rightarrow}\left\vert
\frac{\partial\theta}{\partial\theta^{\prime}}\right\vert d^{n}\theta^{\prime
}\text{,} \label{pre1}%
\end{equation}
respectively. Therefore, it follows that%
\begin{equation}
\sqrt{g\left(  \theta\right)  }d^{n}\theta\overset{\theta\rightarrow
\theta^{\prime}}{\rightarrow}\sqrt{g\left(  \theta^{\prime}\right)  }%
d^{n}\theta^{\prime}\text{.} \label{pre3}%
\end{equation}
Equation (\ref{pre3}) implies that the infinitesimal statistical volume
element is invariant under general coordinate transformations that preserve
orientation (that is, with positive Jacobian). For more details on these
aspects, we suggest Caticha's $2012$ tutorial \cite{tutorial}.

\subsection{Information geometric complexity}

The elements (or points) $\left\{  p\left(  x|\theta\right)  \right\}  $ of an
$n$-dimensional curved statistical manifold $\mathcal{M}_{s}$ are parametrized
using $n$ real valued variables $\left(  \theta^{1}\text{,..., }\theta
^{n}\right)  $,%
\begin{equation}
\mathcal{M}_{s}\overset{\text{def}}{=}\left\{  p\left(  x|\theta\right)
:\theta=\left(  \theta^{1}\text{,..., }\theta^{n}\right)  \in\mathcal{D}%
_{\theta}^{\left(  \text{tot}\right)  }\right\}  \text{.}%
\end{equation}
The set $\mathcal{D}_{\theta}^{\left(  \text{tot}\right)  }$ is the entire
parameter space (available to the system) and is a subset of $%
\mathbb{R}
^{n}$,%
\begin{equation}
\mathcal{D}_{\theta}^{\left(  \text{tot}\right)  }\overset{\text{def}}{=}%
{\displaystyle\bigotimes\limits_{k=1}^{n}}
\mathcal{I}_{\theta^{k}}=\left(  \mathcal{I}_{\theta^{1}}\otimes
\mathcal{I}_{\theta^{2}}\text{...}\otimes\mathcal{I}_{\theta^{n}}\right)
\subseteq%
\mathbb{R}
^{n}%
\end{equation}
where $\mathcal{I}_{\theta^{k}}$ is a subset of $%
\mathbb{R}
$ and represents the entire range of allowable values for the macrovariable
$\theta^{k}$. For example, considering the statistical manifold of
one-dimensional Gaussian probability distributions parametrized in terms of
$\theta=\left(  \mu\text{, }\sigma\right)  $, we obtain%
\begin{equation}
\mathcal{D}_{\theta}^{\left(  \text{tot}\right)  }\overset{\text{def}}%
{=}\mathcal{I}_{\mu}\otimes\mathcal{I}_{\sigma}=\left[  \left(  -\infty\text{,
}+\infty\right)  \otimes\left(  0\text{, }+\infty\right)  \right]  \text{,}%
\end{equation}
with $\mathcal{I}_{\mu}\otimes\mathcal{I}_{\sigma}\subseteq%
\mathbb{R}
^{2}$. In the IGAC, we are interested in a probabilistic description of the
evolution of a given system in terms of its corresponding probability
distribution on $\mathcal{M}_{s}$ which is homeomorphic to $\mathcal{D}%
_{\Theta}^{\left(  \text{tot}\right)  }$. Assume we are interested in the
evolution from $\tau_{\text{initial}}$ to $\tau_{\text{final}}$. Within the
present probabilistic description, this is equivalent to studying the shortest
path (or, in terms of the MrE methods \cite{caticha-giffin}, the maximally
probable path) leading from $\theta\left(  \tau_{\text{initial}}\right)  $ to
$\theta\left(  \tau_{\text{final}}\right)  $.

Is there a way to quantify the complexity of such path? We propose the
so-called information geometric entropy (IGE) $\mathcal{S}_{\mathcal{M}_{s}%
}\left(  \tau\right)  $ as a good complexity quantifier \cite{carlo-PD}. In
what follows, we highlight the key-points leading to the construction of this quantity.

The IGE, an indicator of temporal complexity of geodesic paths within the IGAC
framework, is defined as \cite{carlo-PD},%
\begin{equation}
\mathcal{S}_{\mathcal{M}_{s}}\left(  \tau\right)  \overset{\text{def}}{=}%
\log\widetilde{vol}\left[  \mathcal{D}_{\theta}\left(  \tau\right)  \right]
\text{,}%
\end{equation}
where the average dynamical statistical volume\textbf{\ }$\widetilde
{vol}\left[  \mathcal{D}_{\theta}\left(  \tau\right)  \right]  $%
\textbf{\ }(which we also choose to name the information geometric complexity
(IGC)) is given by,%
\begin{equation}
\widetilde{vol}\left[  \mathcal{D}_{\theta}\left(  \tau\right)  \right]
\overset{\text{def}}{=}\frac{1}{\tau}\int_{0}^{\tau}d\tau^{\prime}vol\left[
\mathcal{D}_{\theta}\left(  \tau^{\prime}\right)  \right]  \text{.}
\label{rhs}%
\end{equation}
Note that the tilde symbol in (\ref{rhs}) denotes the operation of temporal
average. The volume\textbf{\ }$vol\left[  \mathcal{D}_{\theta}\left(
\tau^{\prime}\right)  \right]  $\textbf{\ }in the RHS of (\ref{rhs}) is given
by,%
\begin{equation}
vol\left[  \mathcal{D}_{\theta}\left(  \tau^{\prime}\right)  \right]
\overset{\text{def}}{=}\int_{\mathcal{D}_{\theta}\left(  \tau^{\prime}\right)
}\rho_{\left(  \mathcal{M}_{s}\text{, }g\right)  }\left(  \theta
^{1}\text{,..., }\theta^{n}\right)  d^{n}\theta\text{,} \label{v}%
\end{equation}
where $\rho_{\left(  \mathcal{M}_{s}\text{, }g\right)  }\left(  \theta
^{1}\text{,..., }\theta^{n}\right)  $ is the so-called Fisher density and
equals the square root of the determinant of the metric tensor $g_{\mu\nu
}\left(  \theta\right)  $ with $\theta\equiv\left(  \theta^{1}\text{,...,
}\theta^{n}\right)  $,%
\begin{equation}
\rho_{\left(  \mathcal{M}_{s}\text{, }g\right)  }\left(  \theta^{1}\text{,...,
}\theta^{n}\right)  \overset{\text{def}}{=}\sqrt{g\left(  \theta\right)
}\text{.}%
\end{equation}
The integration space $\mathcal{D}_{\theta}\left(  \tau^{\prime}\right)  $ in
(\ref{v}) is defined as follows,%
\begin{equation}
\mathcal{D}_{\theta}\left(  \tau^{\prime}\right)  \overset{\text{def}}%
{=}\left\{  \theta:\theta^{k}\left(  0\right)  \leq\theta^{k}\leq\theta
^{k}\left(  \tau^{\prime}\right)  \right\}  \text{,} \label{is}%
\end{equation}
where $k=1$,.., $n$ and $\theta^{k}\equiv\theta^{k}\left(  s\right)  $ with
$0\leq s\leq\tau^{\prime}$ such that,%
\begin{equation}
\frac{d^{2}\theta^{k}}{ds^{2}}+\Gamma_{lm}^{k}\frac{d\theta^{l}}{ds}%
\frac{d\theta^{m}}{ds}=0\text{.}%
\end{equation}
The integration space $\mathcal{D}_{\theta}\left(  \tau^{\prime}\right)  $ in
(\ref{is}) is an $n$-dimensional subspace of the whole (permitted) parameter
space $\mathcal{D}_{\theta}^{\left(  \text{tot}\right)  }$. The elements of
$\mathcal{D}_{\theta}\left(  \tau^{\prime}\right)  $ are the $n$-dimensional
macrovariables $\left\{  \theta\right\}  $ whose components $\theta^{k}$ are
bounded by specified limits of integration $\theta^{k}\left(  0\right)  $ and
$\theta^{k}\left(  \tau^{\prime}\right)  $ with $k=1$,.., $n$. The limits of
integration are obtained via integration of the $n$-dimensional set of coupled
nonlinear second order ordinary differential equations characterizing the
geodesic equations. Formally, the IGE is defined in terms of a averaged
parametric $\left(  n+1\right)  $-fold integral ($\tau$ is the parameter) over
the multidimensional geodesic paths connecting $\theta\left(  0\right)  $ to
$\theta\left(  \tau\right)  $. Further conceptual details about the IGE and
the IGC can be found in \cite{carlo-AMC}.

\section{Applications}

In the following, we outline several selected applications concerning the
complexity characterization of geodesic paths on curved statistical manifolds
within the IGAC framework.

\subsection{\textbf{Gaussian statistical models}}

In \cite{carlo-PD, carlo-IJTP}, we apply the IGAC to study the dynamics of a
system with $l$ degrees of freedom, each one described by two pieces of
relevant information, its mean expected value and its variance (Gaussian
statistical macrostates). This leads to consider a statistical model on a
non-maximally symmetric $2l$-dimensional statistical manifold $\mathcal{M}%
_{s}$. It is shown that $\mathcal{M}_{s}$ possesses a constant negative scalar
curvature proportional to the number of degrees of freedom of the system,
$\mathcal{R}_{\mathcal{M}_{s}}=-l$. It is found that the system explores
statistical volume elements on $\mathcal{M}_{s}$ at an exponential rate. The
information geometric entropy $\mathcal{S}_{\mathcal{M}_{s}}$\ increases
linearly in time (statistical evolution parameter) and, moreover, is
proportional to the number of degrees of freedom of the system, $\mathcal{S}%
_{\mathcal{M}_{s}}$ $\overset{\tau\rightarrow\infty}{\sim}l\lambda\tau$ where
$\lambda$ is the maximum positive Lyapunov exponent characterizing the model.
The geodesics on $\mathcal{M}_{s}$ are hyperbolic trajectories. Using the
Jacobi-Levi-Civita (JLC) equation for geodesic spread, we show that the Jacobi
vector field intensity $J_{\mathcal{M}_{s}}$ diverges exponentially and is
proportional to the number of degrees of freedom of the system,
$J_{\mathcal{M}_{s}}$ $\overset{\tau\rightarrow\infty}{\sim}l\exp\left(
\lambda\tau\right)  $. The exponential divergence of the Jacobi vector field
intensity $J_{\mathcal{M}_{s}}$ is a \textit{classical} feature of chaos.
Therefore, we conclude \ that $\mathcal{R}_{\mathcal{M}_{s}}=-l$,
$J_{\mathcal{M}_{s}}\overset{\tau\rightarrow\infty}{\sim}l\exp\left(
\lambda\tau\right)  $ and $\mathcal{S}_{\mathcal{M}_{s}}\overset
{\tau\rightarrow\infty}{\sim}l\lambda\tau$. Thus, $\mathcal{R}_{\mathcal{M}%
_{s}}$, $\mathcal{S}_{\mathcal{M}_{s}}$ and $J_{\mathcal{M}_{s}}$ behave as
proper indicators of chaoticity and are proportional to the number of
Gaussian-distributed microstates of the system. This proportionality, even
though proven in a very special case, leads to conclude there may be a
substantial link among these information geometric indicators of chaoticity.

\subsection{Gaussian statistical models and correlations}

In \cite{carloPA2010}, we apply the IGAC to study the information constrained
dynamics of a system with $l=2$ microscopic degrees of freedom. As working
hypothesis, we assume that such degrees of freedom are represented by two
correlated Gaussian-distributed microvariables characterized by the same
variance. We show that the presence of microcorrelations lead to the emergence
of an asymptotic information geometric compression of the statistical
macrostates explored by the system at a faster rate than that observed in
absence of microcorrelations. This result constitutes an important and
explicit connection between micro-correlations and macro-complexity in
statistical dynamical systems. The relevance of our finding is twofold: first,
it provides a neat description of the effect of information encoded in
microscopic variables on experimentally observable quantities defined in terms
of dynamical macroscopic variables; second, it clearly shows the change in
behavior of the macroscopic complexity of a statistical model caused by the
existence of correlations at the underlying microscopic level.

\subsection{Random frequency macroscopic IHOs}

\emph{\ }The problem of General Relativity is twofold: one is how geometry
evolves, and the other is how particles move in a given geometry. The IGAC
focuses on how particles move in a given geometry and neglects the other
problem, the evolution of the geometry. The realization that there exist two
separate and distinct problems was a turning point in our research and lead to
an unexpected result. In \cite{caticha-cafaro}, we explore the possibility of
using well established principles of inference to derive Newtonian dynamics
from relevant prior information codified into an appropriate statistical
manifold. The basic assumption is that there is an irreducible uncertainty in
the location of particles so that the state of a particle is defined by a
probability distribution. The corresponding configuration space is a
statistical manifold the geometry of which is defined by the Fisher-Rao
information metric. The trajectory follows from a principle of inference, the
MrE method. There is no need for additional physical postulates such as an
action principle or equation of motion, nor for the concept of mass, momentum
and of phase space, not even the notion of time. The resulting entropic
dynamics reproduces Newton's mechanics for any number of particles interacting
among themselves and with external fields. Both the mass of the particles and
their interactions are explained as a consequence of the underlying
statistical manifold.

Following this line of reasoning, in \cite{carlo-CSF, EJTP} we present an
information geometric analogue of the Zurek-Paz quantum chaos criterion in the
\textit{classical reversible limit}. This analogy is illustrated by applying
the IGAC to a set of\textbf{\ }$n$\textbf{-}uncoupled three-dimensional
anisotropic inverted harmonic oscillators (IHOs) characterized by a Ohmic
distributed frequency spectrum.

\subsection{Regular and chaotic quantum spin chains}

In \cite{cafaroMPLB, cafaroPA}, we study the entropic dynamics on curved
statistical manifolds induced by classical probability distributions of common
use in the study of regular and chaotic quantum energy level statistics.
Specifically, we propose an information geometric characterization of chaotic
(integrable) energy level statistics of a quantum antiferromagnetic Ising spin
chain in a tilted (transverse) external magnetic field. We consider the IGAC
of a Poisson distribution coupled to an Exponential bath (spin chain in a
\textit{transverse} magnetic field, regular case) and that of a Wigner-Dyson
distribution coupled to a Gaussian bath (spin chain in a \textit{tilted}
magnetic field, chaotic case). Remarkably, we show that in the former case the
IGE exhibits asymptotic logarithmic growth while in the latter case the IGE
exhibits asymptotic linear growth. In view of these findings, we conjecture
our IGAC might find some potential physical applications in quantum energy
level statistics as well.

\subsection{Complexity reduction and statistical embedding}

In \cite{cafaro-mancini}, we characterize the complexity of geodesic paths on
a curved statistical manifold $\mathcal{M}_{s}$ through the asymptotic
computation of the\ IGC and the Jacobi vector field intensity $J_{\mathcal{M}%
_{s}}$. The manifold $\mathcal{M}_{s}$ is a $2l$-dimensional Gaussian model
reproduced by an appropriate embedding in a larger $4l$-dimensional Gaussian
manifold and endowed with a Fisher-Rao information metric $g_{\mu\nu}\left(
\theta\right)  $ with non-trivial off diagonal terms. These terms emerge due
to the presence of a correlational structure (embedding constraints) among the
statistical variables on the larger manifold and are characterized by
macroscopic correlational coefficients $r_{k}$.\ First, we observe a power law
decay of the information geometric complexity at a rate determined by the
coefficients $r_{k}$ and conclude that the non-trivial off diagonal terms lead
to the emergence of an asymptotic information geometric compression of the
explored macrostates on $\mathcal{M}_{s}$. Finally,\ we also observe that the
presence of such embedding constraints leads to an attenuation of the
asymptotic exponential divergence of the Jacobi vector field intensity. We are
confident the work presented in \cite{cafaro-mancini} constitutes a further
non-trivial step towards the characterization of the complexity of
microscopically correlated multidimensional Gaussian statistical models, and
other models of relevance in realistic physical systems.

\subsection{Scattering induced quantum entanglement}

In \cite{kim1, kim2}, we present an information geometric analysis of
entanglement generated by\textbf{\ }$s$\textbf{-}wave scattering between two
Gaussian wave packets. We conjecture that the pre and post-collisional quantum
dynamical scenarios related to an elastic head-on collision are macroscopic
manifestations emerging from microscopic statistical structures. We
then\textbf{\ }describe them by uncorrelated and correlated Gaussian
statistical models, respectively. This allows us to express the entanglement
strength in terms of scattering potential and incident particle energies.
Furthermore, we show how the entanglement duration can be related\textbf{\ }to
the scattering potential and incident particle energies. Finally, we discuss
the connection between entanglement and complexity of motion. We are confident
that the work presented in \cite{kim1, kim2} represents significant progress
toward the goal of understanding the relationship between statistical
microcorrelations and quantum entanglement on the one hand and the effect of
microcorrelations on the complexity of informational geodesic flows on the
other. It is also our hope to build upon the techniques employed in this work
to ultimately establish a sound information geometric interpretation of
quantum entanglement together with its connection to complexity of motion in
more general physical scenarios.

\subsection{Suppression of classical chaos and quantization}

In \cite{OSID}, we study the information geometry and the entropic dynamics of
a $3d$ Gaussian statistical model. We then compare our analysis to that of a
$2d$ Gaussian statistical model obtained from the higher-dimensional model via
introduction of an additional information constraint that resembles the
quantum mechanical canonical minimum uncertainty relation. We show that the
chaoticity (temporal complexity) of the $2d$ Gaussian statistical model,
quantified by means of the IGE and the Jacobi vector field intensity, is
softened with respect to the chaoticity of the $3d$ Gaussian statistical
model. In view of the similarity between the information constraint on the
variances and the phase-space coarse-graining imposed by the Heisenberg
uncertainty relations, we suggest that our work provides a possible way of
explaining the phenomenon of suppression of classical chaos operated by quantization.

In the same vein of our work in \cite{OSID}, a recent investigation claims
that quantum mechanics can reduce the statistical complexity of classical
models \cite{nc}. Specifically, it was shown that mathematical models
featuring quantum effects can be as predictive as classical models although
implemented by simulators that require less memory, that is, less statistical
complexity. Of course, these two works use different definitions of complexity
and their ultimate goal is definitively not the same. However, it is
remarkable that both of them exploit some quantum feature, Heisenberg's
uncertainty principle in \cite{OSID} and the quantum state discrimination
(information storage) method in \cite{nc}, to exhibit the complexity softening effects.

Is there any link between Heisenberg's uncertainty principle and quantum state
discrimination? Recently, it was shown that any violation of uncertainty
relations in quantum mechanics also leads to a violation of the second law of
thermodynamics \cite{ester}. In addition, it was reported in \cite{ralph} that
a violation of Heisenberg's uncertainty principle allows perfect state
discrimination of nonorthogonal states which, in turn, violates the second law
of thermodynamics \cite{peres}. The possibility of distinguishing
nonorthogonal states is directly related to the question of how much
information we can store in a quantum state. Information storage and memory
are key quantities for the characterization of statistical complexity. In view
of these considerations, it would be worthwhile exploring the possible
thermodynamic link underlying these two different complexity measures.

\section{Closing Remarks}

In this Contribution, we presented our information geometric measure of
complexity of geodesic paths on curved statistical manifolds underlying the
entropic dynamics of classical physical systems described by probability
distributions within the IGAC framework. We also provided several illustrative
examples of entropic dynamical models used to infer macroscopic predictions
when only partial knowledge of the microscopic nature of the system is
available. Finally, among other things, we also presented entropic arguments
to briefly address complexity softening effects due to statistical embedding procedures.

All too often that which is correct is not new and that which is new is not
correct. Being moderately conservative people, we hope that what we presented
satisfies at least of one these two sub-optimal situations. We are aware that
several issues remain unsolved within the IGAC framework and much more work
remains to be done. However, we are immensely gratified\textbf{\ }that our
scientific vision is gaining more attention and is becoming a source of
inspiration for other researchers \cite{peng}.

To conclude, we would like to outline the three possible lines of research for
future investigations:

\begin{itemize}
\item Extend the IGAC to a fully quantum setting where density matrices play
the analogous role of the classical probability distributions: since quantum
computation can be viewed as geometry \cite{mike, army} and computational
tasks have, in general, a thermodynamic cost \cite{charlie}, we might envision
a \emph{thermodynamics of quantum information geometric flows on manifolds of
density operators} whose ultimate internal consistency check forbids the
prediction of the impossible thermodynamic machine.

\item Understand the role of thermodynamics as the possible bridge among
different complexity measures: softening effects in the classical-to-quantum
transitions can occur provided that the various quantum effects being
exploited by the different complexity measures do not violate the second law
of thermodynamics;

\item Describe and understand the role of thermodynamics within the IGAC:
thermodynamics plays a prominent role in the entropic analysis of chaotic
dynamics \cite{beck}. Chaoticity and entropic arguments are the bread and
butter of the IGAC. Furthermore, inspired by \cite{tonyb}, we could
investigate the possible connection between thermodynamics inefficiency
measured by dissipation and ineffectiveness of entropic dynamical models in
making reliable macroscopic predictions.
\end{itemize}

\begin{acknowledgments}
I acknowledge that this work reflects my academic interaction with the
following scientists: Erik Bollt (Potsdam-NY, USA), Carlo Bradaschia (Pisa,
Italy), \emph{Ariel Caticha} (Albany-NY, USA), Giancarlo Cella (Pisa, Italy),
Stefano Mancini (Camerino, Italy), Jie Sun (Potsdam-NY, USA), Peter van Loock
(Erlangen and Mainz, Germany). I also thank Sean Alan Ali and Adom Giffin
whose careful reviews strengthened this Contribution and Denis Gonta for
technical assistance. Finally, special thanks go to Gianpaolo Beretta for
inviting me as an Invited Speaker at \textit{The 12th Biannual Joint European
Thermodynamics Conference} in Brescia, Italy. During this beautifully
organized Conference, I enjoyed talking to Gian Paolo Beretta (Italy), Thomas
Loimer (Austria), Jan Naudts (Belgium), Robert Niven (Australia), Wilhelm
Schneider (Austria), and Qiang Yang (China).
\end{acknowledgments}

\end{document}